\documentclass[aps,prd,longbibliography,preprintnumbers,
nofootinbib,showpacs]{revtex4-2}

\usepackage[utf8]{inputenc}
\usepackage{graphicx}
\usepackage{dcolumn}
\usepackage{amsmath}
\usepackage{amssymb}
\usepackage{amsthm}
\usepackage[framemethod=tikz]{mdframed}
\usepackage{float}
\usepackage{vmargin}
\definecolor{lcolor}{rgb}{0.5,0,0}
\definecolor{citcolor}{rgb}{0,0,1}
\usepackage[breaklinks,colorlinks,urlcolor=blue,citecolor=citcolor,linkcolor=lcolor,linktoc=all]{hyperref}
\usepackage{orcidlink}

\mdfdefinestyle{simplestyle}{
  hidealllines=true,
  leftline=true,
  innerleftmargin=10pt,
  innerrightmargin=10pt,
  innertopmargin=0pt,
}
\theoremstyle{remark}
\newtheorem*{ex}{QM Example}
\surroundwithmdframed[style=simplestyle]{ex}

\renewcommand{\div}{\mathrm{div}}

\newcommand{\chiinv}[2]{\left(\chi^{-1}\right)_{#1}^{\ #2}}

\newcommand{\Tmat}{\Omega^{-1}}
\newcommand{\Tmati}{\Omega}
\newcommand{\set}[1]{\lbrace #1 \rbrace}
\newcommand{\muhat}{\boldsymbol{\hat{\mu}}}
\newcommand{\bvec}{\boldsymbol{z}}
\newcommand{\tr}{\mathrm{Tr}}
\renewcommand{\det}{\mathrm{Det}}

\allowdisplaybreaks    

\begin{document}

\title{Bulk viscosity and $n$-component fluids}
\author{Saga Säppi
\orcidlink{0000-0002-2920-8038}}
\email{sappi@ieec.cat}
\affiliation{Institute of Space Sciences (ICE, CSIC), Campus UAB,  Carrer de Can Magrans, 08193 Barcelona, Spain}
\affiliation{Institut d'Estudis Espacials de Catalunya (IEEC), 08034 Barcelona, Spain}

\begin{abstract}
Understanding the hydrodynamics of out-of-equilibrium dense viscous fluids is of key importance to accurate descriptions of physical systems such as compact stars, particularly their mergers. We consider a near-equilibrium relativistic fluid with $n$ independent and small chemical potentials restoring the system back towards equilibrium. By diagonalising the evolution equation for the out-of-equilibrium chemical potentials, we construct an explicit evolution equation for the bulk scalar of the system in second-order hydrodynamics in terms of equilibrium quantities. We find expressions for the $2n$ transport quantities and show that in the rest frame of the fluid, the system admits a Green's function corresponding to an $n$-component fluid.
\end{abstract}

\maketitle

\section{Introduction} \label{sec:intro}

There are many open questions in the field of transport phenomena in dense relativistic fluids, with many existing studies focussing on a specific equation of state for the fluid. Such systems are found in nature, with a prominent example being neutron star mergers observed through their gravitational wave signal \cite{LIGOScientific:2017ync,LIGOScientific:2017vwq}. Neutron star mergers provide an environment where a dense near-equilibrium fluid is being brought towards equilibrium by chemical reactions. Accurate descriptions of the underlying phenomena is one of the key steps towards  a proper understanding dense, strongly interacting systems, and especially transport in such matter \cite{Baiotti:2016qnr}.

 In this manuscript, we are particularly interested in the case where a repeated compression--decompression cycle causes energy dissipation due to a finite bulk viscosity, which consequently has an effect on the observed merger signal \cite{Alford:2017rxf,Most:2022yhe,Alford:2019oge,Chabanov:2023blf,Chabanov:2023abq}. The qualitative behaviour of the bulk viscosity is known to be different for different types of matter present in neutron stars, namely, for hadronic matter and quark matter \cite{CruzRojas:2024etx,Hernandez:2024rxi,Jyothilakshmi:2022hys,Alford:2022ufz,Alford:2019kdw}---the transport properties depend on the equation of state of the system as well as the rates of the relevant chemical reactions.

With a simple system involving few particle species, the relevant equations derived using relativistic hydrodynamics are largely manageable, but quickly become unwieldy for a more complicated system if treated directly \cite{CruzRojas:2024etx,Hernandez:2024rxi,Harris:2024ssp}. A large number of particle species appears in particular in the hadronic phase, where the number of potentially active composite particles can be significant. 

In this manuscript, we will be rather general: Instead of selecting a specific equation of state---or even necessarily one associated with neutron stars---we consider a general dense fluid with $n$ independent chemical potentials associated with reactions bringing the system towards equilibrium.  By doing so, we provide a general framework for analysing relativistic bulk viscous fluids in the linearised regime irrespective of the number of particle species. We derive the relevant evolution equation for such systems, and show that under certain general conditions they factorise into $n$ subsystems. We also find expressions for the associated transport coefficients in terms of equilibrium quantities. The derivation as well as the equations themselves are useful not only for a better grasp of the general behaviour of relativistic dense viscous fluids, but also for describing complicated systems where $n$ is larger than one or two, important for a more nuanced understanding of neutron star merger dynamics.  

The paper is structured as follows: In Section \ref{sec:defns} we first define precisely what characterises our near-equilibrium system. This includes writing down a system of evolution equations for the set of chemical potentials associated with the chemical reactions bringing the system towards equilibrium. Next, in Section \ref{sec:diag} we diagonalise the set of equations to bring it into a solvable form. In \ref{sec:transport} we write down the equation for the bulk scalar, finding the associated Green's function as well as the relevant transport coefficients in terms of equilibrium quantities. As the notation used can be somewhat abstract, in each of the Sections \ref{sec:defns}--\ref{sec:transport} we complement the definitions and derivations with a corresponding example for $n=2$, associated with (unpaired) quark matter (QM) in the absence of neutrinos (and based on \cite{Hernandez:2025zxw}). We emphasise that this is purely for illustrative purposes, and the treatment in general is applicable for a much wider class of systems. Lastly, in \ref{sec:discussion} we discuss the results.

\section{Linearised $n$-component fluid} \label{sec:defns}

In this section, we outline a general formulation for the relevant physical system. Consider a system of $k>1$ particle species $\set{p_a}_{a\in K}, |K|=k$ with associated chemical potentials $\set{ \mu_{a} }_{a \in K}$ whose chemical equilibrium has been disturbed and is being restored by a set of chemical reactions $\mathcal{C}$. The disturbance of equilibrium manifests as the chemical potentials becoming dependent on a spacetime point. Generally, for some $C\in\mathcal{C}$ the chemical reactions take the form $\sum_{a \in K} x^{C}_a p_{a} \rightarrow \sum_{b \in K} \tilde{x}^{C}_b p_{b}$ (they also come with a corresponding set of reverse reactions $C'\in\mathcal{C'}$ of the form $\sum_{a \in K} x^{C'}_a p_{a} \leftarrow \sum_{b \in  K} \tilde{x}^{C'}_b p_{b}$ ) for some integer coefficients\footnote{Often, these coefficients are all equal to either one or zero.} $x^{C},\tilde{x}^{C}$ ($x^{C'},\tilde{x}^{C'}$), which we set to zero for particle species not appearing in the reaction.

\begin{ex}
 $K=\set{u,d,s,e(,\nu)}$ and the relevant reactions are 
\begin{equation}
	C_1: u+s \longleftrightarrow u+d,\quad C_2: u+e \longrightarrow d+(\nu),\quad C_3: u+e \longrightarrow s+(\nu),
\end{equation}
where we suppress neutrinos $\nu$ from now on for simplicity, so that $k=4$. Thus,
\begin{align}
	\set{x^{C_1}_u,x^{C_1}_d,x^{C_1}_s,x^{C_1}_e}&=\set{1,0,1,0},
\quad \set{\tilde{x}^{C_1}_u,\tilde{x}^{C_1}_d,\tilde{x}^{C_1}_s,\tilde{x}^{C_1}_e}=\set{1,1,0,0}; \nonumber \\	
	\quad\set{x^{C_2}_u,x^{C_2}_d,x^{C_2}_s,x^{C_2}_e}&=\set{1,0,0,1},\quad\set{\tilde{x}^{C_2}_u,\tilde{x}^{C_2}_d,\tilde{x}^{C_2}_s,\tilde{x}^{C_2}_e}=\set{0,1,0,0}; \nonumber \\
	\set{x^{C_3}_u,x^{C_3}_d,x^{C_3}_s,x^{C_3}_e}&=\set{1,0,0,1}, \quad\set{\tilde{x}^{C_3}_u,\tilde{x}^{C_3}_d,\tilde{x}^{C_3}_s,\tilde{x}^{C_3}_e}=\set{0,0,1,0}.
\end{align}
\end{ex}
The system becomes simple to treat if we assume it to be near equilibrium. That is to say, we assume that the  reaction rates of the equation $\Gamma$ can be linearised and written as $\Gamma_C-\Gamma_{C'}\equiv \hat{\mu}_C\lambda_C$ for any $C\in\mathcal{C}$ (and associated $C'\in\mathcal{C'}$), for small \emph{reaction chemical potentials} $\hat{\mu}_C$ describing deviation from chemical equilibrium and a corresponding linearised rate $\lambda_C$. Now, each $\hat{\mu}_C$ can be written as a linear combination of the $\set{\mu_a}_{a\in K}$:
\begin{equation}
\hat{\mu}_C=\sum_{a \in K} \left(x^{C}_a-\tilde{x}^C_b\right) \mu_{a}.
\label{eq:muhatdef}
\end{equation}
For equations involving chemical potentials and related linearised quantities (notably particle densities), we always drop the remainder terms for notational simplicity; everything is derived assuming linearised chemical potentials. To be more precise, the reaction chemical potentials are small in the sense that we require\footnote{It is often possible to relax this condition by requiring that the reaction chemical potentials $\set{\hat{\mu}_C}_{C\in\mathcal{C}}$ are smaller than only \emph{some} of the chemical potentials associated with the particle species $\set{\mu_a}_{a\in K}$.} that $\hat{\mu}_C / \mu_a \ll 1 \, \forall C\in\mathcal{C},a\in K$. This also implies that the chemical potentials $\set{\mu_a}_{a\in K}$ can be expanded around an equilibrium value. That is to say, $\forall a \in K, \mu_a = \mu_a^{0}+\delta \mu_a$, where $\mu_a^{0}$ corresponds to a value obtained fixing the constraint $\hat\mu_C=0\,\forall C\in\mathcal{C}$, so that in fact in Eq. \eqref{eq:muhatdef} one may replace each $\mu_a$ with $\delta \mu_a$, and $\delta \mu_a/\mu_a^{0} \ll 1 \, \forall{a\in K}$. In practice, the importance of this expansion is that the equilibrium chemical potentials $\set{\mu_a^0}_{a\in K}$ carry no dependence on the spacetime point.
 
 \begin{ex} The linearised reaction chemical potentials and associated rates are
 \begin{align}
 \hat{\mu}_{C_1} \lambda_1 &\equiv  \left(\mu_u + \mu_s - \mu_u -\mu_d\right)\lambda_1 =  \left(\mu_{s}-\mu_{d}\right)\lambda_1, \\
  \hat{\mu}_{C_2} \lambda_2 &\equiv  \left(\mu_u + \mu_e - \mu_d\right)\lambda_2,\\
   \hat{\mu}_{C_3} \lambda_3 &\equiv \left(\mu_u + \mu_e - \mu_s\right)\lambda_3.
 \end{align}
\end{ex}

We take the first $n\leq |\mathcal{C}|=|\mathcal{C'}|$ of the reaction chemical potentials to be independent, in the sense that any possibly remaining $n-|\mathcal{C}|$ rates can be written as 
\begin{equation}
\Gamma_{C_\ell}-\Gamma_{C'_\ell}\equiv \lambda_{C_{\ell}} \sum_{\alpha=1}^{n} r_\alpha^{C_\ell}\hat{\mu}_\alpha,
\label{eq:muhatdef2}
\end{equation}
with $\ell \in \set{
n+1\ldots |\mathcal{C}|} $ (for $1 \leq \ell \leq n$, we define $r_\alpha^{C_\ell}=\delta_{\alpha,C_\ell}$ so that $r_\alpha^{C}$ is well-defined and \eqref{eq:muhatdef2} holds $\forall C\in \mathcal{C}$). We denote $\muhat\equiv \hat{\mu}_\alpha = \set{ \hat{\mu}_1 ,\ldots \hat{\mu}_n}.$ From now on, indices $\alpha,\beta$ will always enumerate these degrees of freedom.

 \begin{ex}  Noting that $\hat{\mu}_{C_3}=\hat{\mu}_{C_2}-\hat{\mu}_{C_1}$, we take the linearly independent chemical potentials to be $\muhat=(\hat{\mu}_1,\hat{\mu}_2)$. Thus, we indeed have $n=2$, and we find $r_{\alpha}^{C_\beta}=\delta_{\alpha}^{\beta}$ for $\alpha,\beta \in \set{1,2}$, as well as $r_{1}^{C_3} =-1,r_{2}^{C_3} =+1$. 
\end{ex}

To make contact with hydrodynamics, consider a fluid with a four-velocity field $u$, and write $\vartheta = \div u=\partial_\nu u^\nu$, $D_u=u_{\nu} \partial^{\nu}$, where the $\set{\partial_\nu}_{\nu=0}^d$ are spacetime derivatives. To fix conventions, we use the mostly-minus metric, although the signature only enters into expressions obtained in fixed frames, and use the summation convention for spacetime indices. In second-order hydrodynamics, $\vartheta$ is formally treated as an independent variable in the evolution equations. Nevertheless, it is associated with the conserved Noether current of particle numbers and connected to them via a continuity equation $ \div n_a u=D_u  n_a +n_a \vartheta=0\, \forall a\in K $, which we will use in the derivations below (and where $\set{n_a}_{a \in K}$ are particle densities defined shortly). 

The pressure of the system in equilibrium $p$ depends on the equilibrium chemical potentials $\set{ \mu_a^0 }_{a \in K}$, the equilibrium temperature $T$, and possible other variables that are independent of the spacetime point, such as particle masses. The bulk scalar of the system $\Pi=p_{\mathrm{non-eq.}}-p$ measures deviation from equilibrium in terms of the non-equilibrium pressure $p_{\mathrm{non-eq.}}$, and takes the form $\Pi=\sum_{\alpha=1}^{n} \Pi_\alpha \hat{\mu}_\alpha\equiv \langle \boldsymbol{\Pi},\muhat\rangle$ in the near-equilibrium linearised approximation. The vector of coefficients $\boldsymbol{\Pi}$ depends on the particle densities $n_a = \partial p / \partial \mu_a^0$ and (generally nondiagonal but symmetric) susceptibilities $\chi_a^{\,b} = \partial^2 p / \partial \mu^0_a \partial \mu^0_b$; $a,b\in K$, all in equilibrium.

Importantly, in the definition of $\Pi$ one may fully incorporate other  constraints such as conservation laws that may exist in the system besides those given by the chemical reactions and the continuity equation. This is vital for treating physical systems, but does not affect the derivation presented here, so the exact form of these constraints and the $\boldsymbol{\Pi}$ is irrelevant for our present purposes. Nevertheless, it is the evolution equation of $\Pi$ we are after.  

\begin{ex} Writing $\delta \mu_a$ for the deviations of the chemical potentials from their equilibrium values, the additional constraints fulfilled in QM can be written in the linearised regime as 
\begin{equation}
\sum_{a\in\set{u,d,s,e}} \left(\chi_{e}^{\,a}-\chi_{u}^{\,a}\right)\delta \mu_a =0 ,\quad\sum_{a\in\set{u,d,s,e}} \left(\chi_{d}^{\,a}+\chi_{s}^{\,a}-\chi_{e}^{\,a}\right)\delta \mu_a =0.
\end{equation}
These correspond to `charge neutrality' and `beta equilibrium' respectively, with their equilibrium counterparts being $\mu_d^0=\mu_s^0=\mu_u^0+\mu_e^0$ and $2n_u-n_d-n_s-3n_e=0$. We note that in this special case these conditions imply that $\mu_u \sim \mu_d \sim \mu_s \ll \mu_e $.  In QM, the two components $\Pi_1,\Pi_2$ of the bulk scalar as well as the equilibrium susceptibilities and densities $\Pi$ depends on can be expressed uniquely in terms of a single equilibrium chemical potential (and the temperature $T$)---for explicit expressions in QM, see the appendix of \cite{Hernandez:2025zxw}. This is not always possible, but it should not be taken as a problem as such, it simply means that the system is characterised by more than one free chemical potential. To reiterate, while the constraints and especially their practical implementation can be somewhat subtle, they are not relevant for the derivation presented here, and only mentioned for completeness for applying the results to physical systems. 
\end{ex}

Following \cite{CruzRojas:2024etx,Hernandez:2025zxw}, we next write down an evolution equation for the reaction chemical potentials. Generally, the near-equilibrium chemical potentials as well as the velocity $u$ itself (or specifically, its divergence $\vartheta$) are taken to be  functions $\mathbb{M}^{d+1} \rightarrow \mathbb{R}$.
 To begin, we note that by definition the rates of the chemical reactions measure the rate of change in the particle densities, and as such the forms described above lead to the following evolution equation:
 \begin{equation}
	D_u \boldsymbol{n}_{\mathrm{non-eq.}} = \lambda \hat{\mu}.
	\label{eq:nevo}
 \end{equation}
 Here, the new (real-valued) quantities $\boldsymbol{n}_{\mathrm{non-eq.}}$ and $\lambda$ are the vector of \emph{non-equilibrium} particle densities whose $a$th component is $\partial p_{\mathrm{non-eq.}} / \partial \mu_a$ and the $k \times n$-matrix of coefficients defined by\footnote{The overall sign is largely conventional, and chosen to agree with \cite{Hernandez:2025zxw} and to counteract the overall sign of $\Tmati$.}, $\lambda_a^{\, \alpha} = -\sum_{C\in\mathcal{C}} (x_a^C-\tilde{x}_a^C) r^{C}_{\alpha} \lambda_{C}$, respectively.

 Next, we note that in the linear regime the expansions $p_{\mathrm{non-eq.}}-p=\sum_{a} n_a \delta\mu_a$, $n_{a,\mathrm{non-eq.}}-n_{a}=\sum_{b}\chi_{a}^{\, b} \delta \mu_b$ hold $\forall a,b\in K$. The third step is to connect the continuity equation to the linearised expansions, and write
\begin{equation}
D_u \boldsymbol{n}_{\mathrm{non-eq.}} = \chi_a^{\, b} D_u \mu_b - \vartheta\boldsymbol{n}_{\mathrm{non-eq.}}.
\label{eq:contevo}
\end{equation} 
We are almost done, but in order to get a convenient form for our purposes, we  isolate the $\muhat$-dependence. To do so we move to a partially rotated basis of the chemical potentials. Starting from the standard basis $\set{ \mu_{a} }_{a\in K}$ we rotate to a basis $\set{ \tilde{\mu}_\mathfrak{a} }_{\mathfrak{a}\in\mathfrak{K}}$ such that for the first $n$ components $\tilde{\mu}$ coincide with each of the $\hat{\mu}$, and for the remaining $|\ell |$ they coincide with some elements of the standard basis chosen so that the rotated basis is linearly independent and $k$-dimensional. To move between these two bases, we define  $M$, a (symmetric) $k\times k$-matrix of inverse susceptibilities. Explicitly, its coefficients read $M_\mathfrak{a}^{\, a} \equiv \chiinv{\mathfrak{a}}{c}$ where for $\mathfrak{a}=b>n$ the elements are simply the inverse susceptibilities $\chiinv{b}{c}$ and for $\mathfrak{a}=\alpha\leq n$ the elements are defined analogously to the corresponding chemical potentials: $\chiinv{\alpha}{c}=\sum_{a \in  K} \left(x^{C\alpha}_a - \tilde{x}^{C_\alpha}_a\right) \chiinv{b}{c}$, for $C_\alpha$ the chemical reaction associated with $\hat{\mu}_\alpha$. This is exactly the matrix appearing when combining Eqs. \eqref{eq:nevo} and \eqref{eq:contevo} and inverting the result in the linearised regime: the resulting evolution equation for the chemical potentials is
\begin{equation}
D_u \boldsymbol{\tilde{\mu}} = M \lambda \muhat - M \vartheta \boldsymbol{n}, \label{eq:odesysfull}
\end{equation}
where $\boldsymbol{n}$ is now a vector of the \emph{equilibrium} particle densities.

In practice, we are only interested in the first $n$ components of Eq. 
\eqref{eq:odesysfull} as those are the ones appearing in $\Pi$.  Writing $\hat{M}$ for the restriction of $M$ to its first $n$ rows (so that $\hat{M}$ is an $n\times k$-matrix), we define the $n\times n$-matrix $\Tmati$ via $\Tmati = -\hat{M}\lambda$, and an auxiliary $n$-dimensional vector\footnote{We do not include $\vartheta$ in the definition of $\bvec$, as in the case relevant for us, we will have $D_u \bvec=0$ but $D_u \vartheta \neq 0$. Note also that $\hat{M},\lambda$ are generally non-square matrices, so that their individual inverses are not defined, and only the inverse of $\Tmati$ is, so that the expression for $\bvec$ cannot generally be simplified further.} $\bvec = \Tmat \hat M \boldsymbol{n}$.

Connecting these quantities to physics, the matrix $\Tmati$ has units of $[\mathrm{time}]^{-1}$ and its eigenvalues will be associated with the (inverse) relaxation times appearing in the fluid description, whereas $\bvec$ has units of $[\mathrm{chemical \, potential}]^2$, and will be associated with the bulk viscous transport coefficients. Note that despite the minus sign appearing in the definition of $\Tmati$, $\lambda$ is constructed such that for the relevant physical systems the eigenvalues of $\Tmati$ are positive.

\begin{ex} The rotated basis is $\set{\mu_1,\mu_2,\mu_s,\mu_e}$, and $\boldsymbol{n}=(n_u,n_d,n_s,n_e)$. The matrix $M$ connects the bases $\set{\mu_u,\mu_d,\mu_s,\mu_e}$ and $\set{\mu_1,\mu_2,\mu_s,\mu_e}$ and reads for $a\in \set{u,d,s,e}$
\begin{equation}
M=\begin{pmatrix}\chiinv{1}{u} & \chiinv{1}{d} & \chiinv{1}{s} & \chiinv{1}{e}\\
\chiinv{2}{u} & \chiinv{2}{d} & \chiinv{2}{s} & \chiinv{2}{e}\\
\chiinv{s}{u} & \chiinv{s}{d} & \chiinv{s}{s} & \chiinv{s}{e}\\
\chiinv{e}{u} & \chiinv{e}{d} & \chiinv{e}{s} & \chiinv{e}{e}
\end{pmatrix},\quad\begin{array}{c}
\chiinv{1}{a}=\chiinv{s}{a}-\chiinv{d}{a}\\
\chiinv{2}{a}=\chiinv{u}{a}+\chiinv{e}{a}-\chiinv{d}{a}
\end{array}.
\end{equation}
Lastly, the matrix of rates is 
\begin{equation}
\lambda=\begin{pmatrix}\lambda_{3} & -\lambda_{2}-\lambda_{3}\\
\lambda_{1} & \lambda_{2}\\
-\lambda_{1}-\lambda_{3} & \lambda_{3}\\
\lambda_{3} & -\lambda_{2}-\lambda_{3}
\end{pmatrix}. 
\end{equation}
The quantities $\Tmati$, $\bvec$ obtained by matrix multiplication do not simplify in a meaningful way in the special case of QM.
\end{ex}

 Considering the first $n$ components, Eq. \eqref{eq:odesysfull} becomes
\begin{equation}
(\Tmat D_u+\mathrm{id}_n ) \muhat = - \bvec \vartheta \iff \Tmati \muhat = - \Tmati \bvec \vartheta - D_u \muhat,  \label{eq:odesys}
\end{equation} 
where both forms will prove useful.  Here (already in the definition of $\bvec$) and from now on, we have implicitly assumed that the matrix $\Tmati$ is invertible. If this is the case, one can factorise the system of equations for $\muhat$. In practice, this is true as long as the rates $\set{ \lambda_{C} }_{C\in\mathcal{C}}$ are all nonvanishing, unless the pressure takes a particularly degenerate form. For a physical system, the degeneracy of $\Tmati$ also essentially implies that one is considering extraneous degrees of freedom: If $\Tmati$ is $r$-fold degenerate, one is in fact dealing with only $n-r$ independent chemical potentials.

\section{Diagonalising the differential equation} \label{sec:diag}

We are now ready to manipulate the system of differential equations. Note that what follows makes \emph{no assumptions} on the specific form or physical meaning of the quantities considered, other than that the square matrices are nondegenerate and the relevant functions otherwise sufficiently well-behaved (that everything is well-defined in the relevant regime of temperatures and chemical potentials and at all spacetime points, in particular that $\vartheta$ is $C^n$-smooth, and eventually that $D_u$ commutes with the coefficient matrix). Thus, the approach is valid for any system of differential equations of the above form.

 To start, we consider the characteristic polynomial of an invertible real-valued $n\times n$-matrix $\Tmati$. To remind the reader, it is given by
 \begin{equation}
  p_\Tmati:\mathbb{R}^{n\times n} \rightarrow \mathbb{R}[x]: \Tmati \mapsto p_{\Tmati}(x)=\sum_{i=0}^n c_i[\Tmati]x^k = \prod_{\omega\in\Sigma[\Tmati]} (x-\omega),  
  \label{eq:charpoldef}
 \end{equation}
where $\Sigma[\Tmati]$ is the set of eigenvalues of $\Tmati$ and the $n+1$ coefficients $\set{c_i[\Tmati]}_{i=0}^n$ are formed of invariants of $\Tmati$ and its powers. For example, $c_{n}\left[\Tmati\right]\equiv1, c_{0}\left[\Tmati\right]\equiv\left(-1\right)^{n}\det \Tmati$. While there exists no simple formula for the $c_{i}$ valid for all $i$, their  exact form is not of particular interest to until we fix $n$, only that they are not degenerate. We denote $\tilde{p}_{\Tmati}\left(\Tmati\right)\Tmati=p_{\Tmati}\left(\Tmati\right)-c_{0}\left[\Tmati\right]\mathrm{id}_{n}$. That is, $\tilde{p}_{\Tmati}\left(\Tmati\right)=\sum_{i=1}^{n}c_{i}\left[\Tmati\right]\Tmati^{i-1}$. Then, by Cayley--Hamilton 
\begin{equation}
\tilde{p}_{\Tmati}\left(\Tmati\right)\Tmati=-c_{0}\left[\Tmati\right]\mathrm{id}_{n}\iff \Tmati^{-1}=-c_{0}^{-1}\left[\Tmati\right]\tilde{p}_{\Tmati}\left(\Tmati\right).
\label{eq:inversecayley}
\end{equation}
\begin{ex}
For $n=2$, we have $c_{0}[\Tmati]=\det\Tmat,c_1[\Tmati]=-\tr \Tmati,c_2[\Tmati]=1$, so that the statement of the Cayley--Hamilton theorem becomes
\begin{equation}
 \Omega^2 - \tr \Tmati \Tmati + \det \Tmati = 0 \iff \Tmat = -\frac{1}{\det \Tmati} \left(\Tmati - \tr \Tmati \mathrm{id}_2\right) = \tr \Tmat \mathrm{id}_2 - \det \Tmat \Tmati. 
 \end{equation}
\end{ex}
Using the representation of Eq. \eqref{eq:inversecayley}, we can write the set of differential equations of the form shown in \eqref{eq:odesys} as a $n$th order differential set of differential equations diagonalised for the $\muhat$, assuming that $[D_u,\Tmati]=0$, which for the physical case holds since $\Tmati$ contains only equilibrium quantities. First, we note that for any $i \geq 1$ 
\begin{equation}
D_{u}\Tmati^{i}\muhat=\Tmati^{i-1}D_{u}\Tmati\muhat=-D_{u}\Tmati^{i}\bvec\vartheta-D_{u}^{2}\Tmati^{i-1}\muhat.
\end{equation}
This recursion gives
\begin{equation}
D_{u}\Tmati^{i}\muhat=\left(-1\right)^{i}D_{u}^{i+1}\muhat+\sum_{j=0}^{i-1}\left(-1\right)^{j+1}D_{u}^{j+1}\Tmati^{i-j}\bvec \vartheta.
\end{equation}
 Substituting this to Eq. \eqref{eq:odesys} by using Eq. \eqref{eq:inversecayley} and extracting the $i=1$-term out of the characteristic polynomial we have
\begin{equation}
\sum_{i=2}^{n}c_{i}\left[\Tmati\right]\left(-1\right)^{i}D_{u}^{i}\muhat-c_{1}\left[\Tmati\right]D_{u}\muhat+c_{0}\left[\Tmati\right]\muhat=-c_{0}\left[\Tmati\right]\bvec\vartheta-\sum_{i=2}^{n}c_{i}\left[\Tmati\right]\sum_{j=0}^{i-2}\left(-1\right)^{j}D_{u}^{j+1}\Tmati^{i-j-1}\bvec\vartheta.
\label{eq:musystemsum}
\end{equation}

Equation \eqref{eq:musystemsum} is already of the desired form. However, we may make it more visually appealing by explicitly extracting the characteristic polynomial. We formally write $\sum_{j=0}^{i-2}\left(-1\right)^{j}\Tmati^{i-j-1}D_{u}^{j+1}=\left(\Tmati+D_{u}\mathrm{id}_{n}\right)^{-1}\left[\Tmati^{i}D_{u}+\left(-1\right)^{i}\Tmati D_{u}^{i}\right]$, which equals zero for $i=1$ and the identity for $i=0$, to obtain 
\begin{align}
\boldsymbol{0}&=\sum_{i=0}^{n}c_{i}\left[\Tmati\right]\left(-D_{u}\right)^{i}\muhat+\frac{1}{D_{u}\mathrm{id}_{n}+\Tmati}\left(\sum_{i=0}^{n}c_{i}\left[\Tmati\right]\left(-D_{u}\right)^{i}\Tmati+D_{u}\overbrace{\sum_{i=0}^{n}c_{i}\left[\Tmati\right]\Tmati^{i}}^{=0}\right)\bvec\vartheta\nonumber \\
&=p_{\Tmati}\left(-D_{u}\right)\left[\muhat+\left(D_{u}\mathrm{id}_n+\Tmati\right)^{-1}\Tmati\bvec\vartheta\right].
\label{eq:musystem}
\end{align}
Above, $p_{\Tmati}(-D_u)$ is to be understood as an $n$th order differential operator whose coefficients are the coefficients of the characteristic polynomial with an alternating sign, and importantly the inverse operator appearing in the equation is nothing nonlocal, but simply a finite (double) sum of the powers of $\Tmati$ and $D_u$ acting on $\bvec \vartheta$. The system of equations is now diagonalised for the $\muhat$, and easy to treat with standard methods. 
 \begin{ex}
When $n=2$, the general formula for the derivatives is not needed, only that   $D_u\Tmati\muhat = -\Tmati \bvec D_u z - \Tmati D_u^2 \muhat$. In addition, Eq. \eqref{eq:musystemsum} is more convenient than  Eq. \eqref{eq:musystem}, reading immediately
\begin{align}
c_2[\Tmati] D_u^2 \muhat - c_1[\Tmati] D_u \muhat + c_0[\Tmati] \muhat &= - c_0 [\Tmati] \bvec \vartheta - c_2 [\Tmati] \Tmati \bvec  D_u \vartheta\nonumber 
\\ \iff D_u^2 \muhat + \tr \Tmati D_u \muhat + \det \Tmati \muhat &= - \det \Tmati \bvec \vartheta - \Omega \bvec D_u \vartheta . 
\end{align}
\end{ex}

\section{Bulk scalar evolution and transport coefficients} \label{sec:transport}

To understand the underlying physical system, we are looking for an evolution equation for the bulk scalar. We contract the vector-valued Eq. \eqref{eq:musystem} with the vector $\boldsymbol{\Pi}$. For reasons of convention and to get the appropriate units, we rescale the equation by multiplying it with $\det \Tmat$; this in particular sets the coefficient of $\Pi$ without derivatives to unity for any $n$. The result is the following differential equation for the bulk scalar (again, we emphasise that it is only the $\set{\muhat_\alpha}_{\alpha=1}^{n}$ and $\vartheta$ that are acted on nontrivially by the derivatives, the components of $\Pi$ appearing in the scalar products are given in terms of equilibrium quantities and commute with the convective derivative): 
\begin{align}
\det \Tmat p_{\Tmati}\left(-D_{u}\right) \Pi &= -  \det \Tmat \langle \Pi , \left(\Tmati+D_{u}\mathrm{id}_{n}\right)^{-1}\Tmati\bvec \rangle \left[ p_{\Tmati}\left(-D_{u}\right)\vartheta \right]\nonumber \\
\implies  \sum_{\alpha=1}^n T_\alpha D_u^\alpha \Pi +\Pi &=- \sum_{\alpha=1}^{n}Z_\alpha D_u^{\alpha-1} \vartheta.
\label{eq:pieq}
\end{align}
Above, we have implicitly defined  $2n$  transport coefficients as $T_\alpha$, the $n$ combinations of \emph{relaxation times} and  $Z_\alpha$, the $n$ \emph{bulk viscous coefficients}. They are all given in terms of equilibrium quantities. While the evolution equation \eqref{eq:pieq} is already of a usable form for e.g. numerical simulations of viscous systems, we can obtain further  qualitative understanding of the behaviour of the corresponding physical systems. Due to the structure of the $\set{c_\alpha(\Tmati)}_{\alpha=1}^n$, the form of the constant terms and the highest derivatives is always fixed: $1$ and $\det \Tmat$ on the left-hand side, $\langle \boldsymbol{\Pi},\bvec \rangle$ and $\det \Tmat \langle \boldsymbol{\Pi},\Tmati \bvec \rangle$  on the right-hand side, respectively. More generally, at any fixed $n\leq 1$ and $i\in\set{1,\ldots,n}$ the coefficient of $D_u^i \Pi$ ($D_u^i \vartheta$) equals the coefficient of $D_u^{i+1} \Pi$ ($D_u^{i+1} \vartheta$) as $n\mapsto n+1$, so that only the coefficient of the first derivative is new at a given order. 

 The equation \eqref{eq:pieq} is an $n$th order linear  differential equation with constant, and by assumption nondegenerate, coefficients. This already implies that the general solution will  be a superposition of exponentials, but to find an explicit solution, we consider the rest frame, so that $D_u \rightarrow \partial_t$, in the linearised approximation (but we treat $\vartheta$ as an independent variable in the spirit of second-order hydrodynamics).  As long as the derivatives $\partial_t^\alpha \vartheta$ are all piecewise continuous on $\mathbb{R}$, the existence of a unique solution to the corresponding initial-value problem is guaranteed, and in particular the system admits a Green's function analysis. As a linear ordinary differential equation with constant coefficients, the form of the Green's function is well known. The homogenous equation admits $n$ linearly independent exponential solutions, and the Green's function is their sum multiplied by the Heaviside function $\theta(t)$. For the physical system under consideration, we look for decaying solutions, so that the general form is
\begin{equation}
G\equiv\sum_\alpha G_\alpha: \mathbb{R}\rightarrow \mathbb{R}\, \mathrm{s.t.}\, t \mapsto G(t)\equiv\theta(t)\sum_{\alpha=1}^{n}\frac{\zeta_\alpha}{\tau_\alpha} e^{-t/\tau_\alpha}.
\label{eq:greensf}
\end{equation}

The physical implication of the form in Eq. \eqref{eq:greensf} is that the system is in a superposition of $n$ separate fluid components, each with a relaxation time $\tau_\alpha$ and a bulk viscosity $\zeta_\alpha$, and each obeying the Israel--Stewart equation for bulk viscosity, the $n=1$ counterpart of the evolution equation \eqref{eq:musystem} $\tau D_{u}\Pi+\Pi+\zeta D_{u}u=0$ for $\tau=\tau_\alpha,\zeta=\zeta_\alpha$. Formally, for each component we then have $\tau=\chi'/\lambda'$ and $\zeta=n'^2/\lambda'$ for some rate, density, and susceptibility $\lambda',n',\chi'$, but these need not correspond to physical quantities.

The general form of \eqref{eq:greensf} is implicit by the structure of the relevant equations. Indeed, just as the case $n=2$ corresponds to the so-called Burgers model in the study of viscoelastic materials, the case for general $n$ corresponds to a generalised Maxwell model \cite{Banks_Hu_Kenz_2011}. In particular, the Green's function of \eqref{eq:greensf} is of the form of a so-called Prony series. In the context of viscoelasticity, this is a commonly used fitting ansatz \cite{park1999methods,schapery1999methods,park2001fitting}. However, here we are able solve the coefficients explicitly in terms of equilibrium quantities.

 \begin{ex} For $n=2$, writing $\tau_+,\tau_-$ for the eigenvalues of $\Tmat$, Eq. \eqref{eq:pieq} becomes
 \begin{align} \tau_+ \tau_- D_u^2 \Pi + \left(\tau_+ +\tau_-\right) D_u \Pi + \Pi  = -\langle \boldsymbol{\Pi},\bvec\rangle \vartheta - \tau_+ \tau_- \langle \boldsymbol{\Pi},\Tmati \bvec\rangle.
 \end{align}
 The Green's function has the form 
 \begin{equation}
 	G(t) = \theta(t)\frac{\zeta_1}{\tau_1} e^{-t/\tau_1} + \theta(t) \frac{\zeta_2}{\tau_2} e^{-t/\tau_2}. 
 \end{equation}
\end{ex}

From the homogeneous equation, we find
\begin{equation}
\det \Tmat p_{\Tmati}\left(\tau_{\alpha}^{-1}\right)G_{\alpha}\left(t\right)=0\,\forall \alpha \in \set{ 1 \ldots n}.
\end{equation}
Given that $\Tmati$ is non-degenerate by assumption, for $G_\alpha$ not identically zero, this dictates that the inverse relaxation times $\set{\tau_\alpha^{-1}}_{\alpha=1}^{n}$ are the eigenvalues of $\Tmati$:
\begin{equation}
T_\alpha = (-1)^{\alpha}\det\Tmat c_\alpha[\Tmati],\quad \set{\tau_\alpha}_{\alpha=1}^{n} = \Sigma[\Tmat]. 
\label{eq:relaxationtimes}
\end{equation}
The latter relation motivates the notation $\Tmat=\mathcal{T}$ in \cite{Hernandez:2025zxw}. For the solutions to be decaying, the relaxation times must be positive and as was noted above, this is the case for physical systems, and we assume so going forward.

\begin{figure}
\centering
    \includegraphics[width=\textwidth]{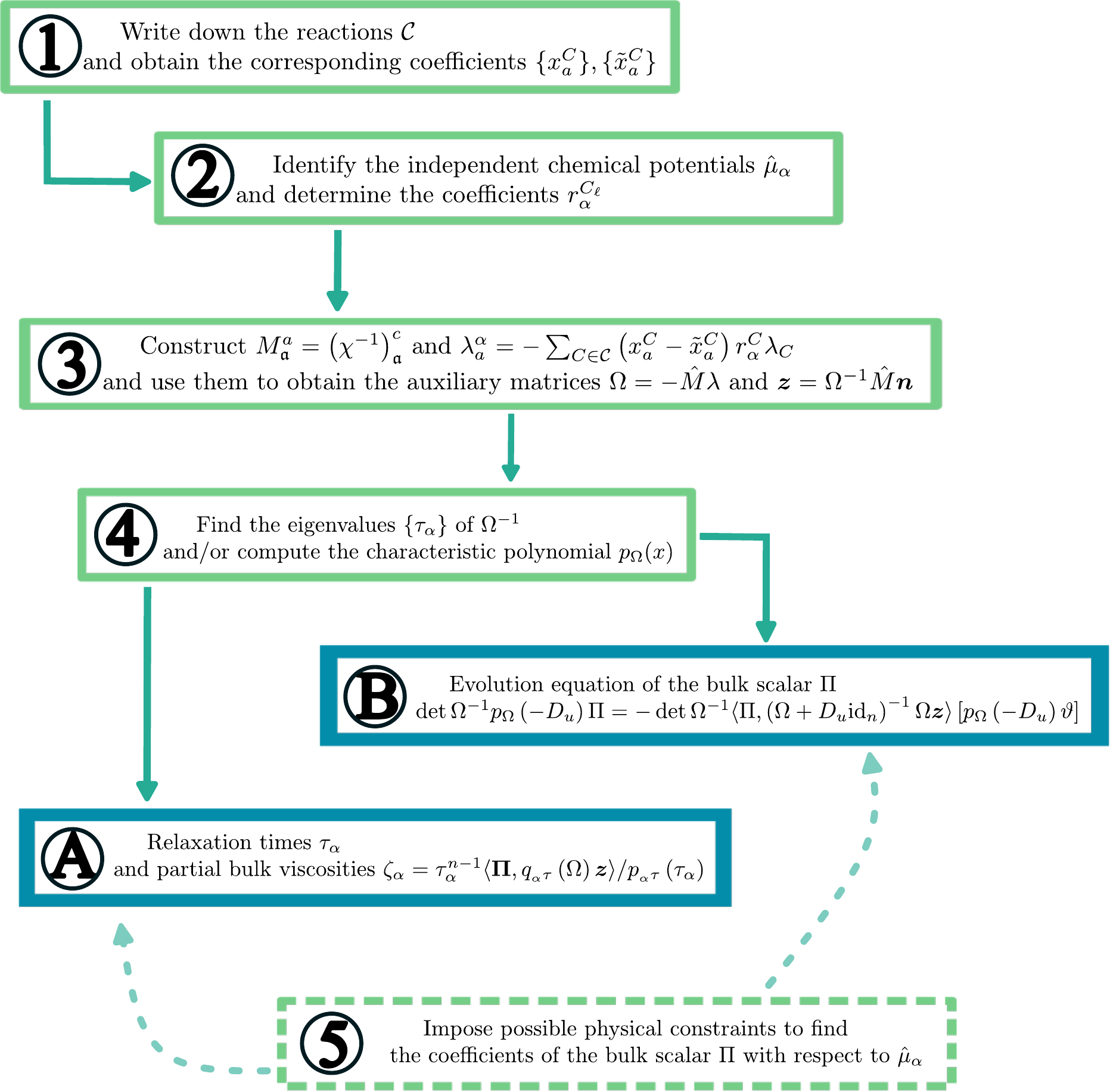}
    \caption{A flowchart demonstrating the steps starting from identifying the appropriate physical system and leading to obtaining the transport coefficients and the evolution equation of the bulk scalar. The details of the fifth step, distinguished by dashed lines, depend on the specifics of the system under consideration.}
    \label{fig:flowchart}
\end{figure}

The coefficients $\set{\zeta_\alpha}_{\alpha=1}^n$ are \emph{partial bulk viscosities}. To relate them to the coefficients $\set{Z_\alpha}_{\alpha=1}^n$, known in terms of equilibrium quantities, we generalise the approach of \cite{Gavassino:2023eoz}. For demonstrative purposes, it suffices if  the bulk scalar $\Pi$ also splits into $n$ distinct components. For instance, if $\vartheta$ varies slowly so that $\vartheta(t-t')\approx\vartheta(t)$ under the convolution integral appearing in the particular solution of the differential equation using the Green's function.\footnote{That is to say, we are in the Navier--Stokes regime.}  The split obtained here is generally distinct from the earlier split of $\Pi=\left\langle \boldsymbol{\Pi},\muhat\right\rangle$, as, denoting $\Pi=\sum_{\alpha=1}^{n}\pi_{\alpha}$,  each component fulfills the $n=1$ equation:
\begin{equation}
\tau_{\alpha}\partial_{t}\pi_{\alpha}+\pi_{\alpha}+\zeta_{\alpha}\vartheta=0\quad\forall\alpha\in\set{1,\ldots, n}.
\end{equation} 
This fixes the partial bulk viscosities $\set{\zeta_\alpha}_{\alpha=1}^n$. We consider 
\begin{equation}
\sum_{\alpha=1}^{n}\prod_{\beta=1,\beta\neq\alpha}^{n}\left(\tau_{\beta}\partial_{t}+1\right)\left(\tau_{\alpha}\partial_{t}\pi_{\alpha}+\pi_{\alpha}+\zeta_{\alpha}\vartheta\right)=0
\end{equation}
and find that the $\pi_\alpha$-dependent parts reduce to Eq. \eqref{eq:pieq} automatically. To match the the $\vartheta$-dependent part, we make some auxiliary definitions. We write $\mathfrak{P}^{\alpha}:\mathbb{R}^{n}\rightarrow\mathbb{R},\ \boldsymbol{y}\mapsto\mathfrak{P}^{\alpha}\left(\boldsymbol{y}\right)$, defined for $i\leq n$, for a function which evaluates to the sum of the products of $\alpha-1$ combinations of the elements of $\boldsymbol{y}$. This quantity is related to terms seen in the characteristic polynomial written in terms of eigenvalues, but without oscillating signs. Indeed, we also define for $m\leq n$ a reciprocal characteristic polynomial as $q:\mathbb{R}^{m \times m }\rightarrow \mathbb{R}[x],A\mapsto q_{A}\left(x\right)=\sum_{i=0}^{m} c_{m-i}\left[A\right]x^i$, and lastly, an $(n-1)\times (n-1)$ matrix $\,_{\beta}\mathcal{T}=\mathrm{diag}\left( \tau_1,\ldots,\tau_{\overline{\beta}},\ldots,\tau_n\right)$, i.e., with the $\beta$th entry removed, and $_{\beta}\boldsymbol{\mathcal{T}}$ for the corresponding $n-1$-vector. Then, we find that (in the  second step substituting the RHS of the first line of Eq. \eqref{eq:pieq}) 
\begin{equation}
Z_\alpha= \sum_{\beta=1}^{n}\mathfrak{P}^{\alpha}\left(_{\beta}\boldsymbol{\mathcal{T}}\right)\zeta_{\beta} \implies \zeta_{\alpha} = \frac{\tau_{\alpha}^{n-1} }{p_{_{\alpha}\mathcal{T}}\left(\tau_{\alpha}\right)}\left\langle \boldsymbol{\Pi},q_{_{\alpha}\mathcal{T}}\left(\Omega\right)\boldsymbol{z}\right\rangle \quad \forall \alpha \in \set{1\ldots n}.
\label{eq:zetasol}
\end{equation}
%
 \begin{ex}
We have $T_1 = \tau_+\tau_-, T_2=\tau_++\tau_-$. The eigenvalues indeed match the relaxation times, $\tau_1 = \tau_+,\tau_2=\tau_-$, and the bulk viscous transport coefficients are 
\begin{align}
Z_1 = \langle \boldsymbol{\Pi},\bvec \rangle,\quad Z_2 &=  \det \Tmat \langle \boldsymbol{\Pi},\Tmati \bvec \rangle \, \nonumber \\ \iff \, \zeta_{1}=\tau_1\frac{\left\langle \boldsymbol{\Pi},\boldsymbol{z}\right\rangle -\tau_{2}\left\langle \boldsymbol{\Pi},\Tmati\boldsymbol{z}\right\rangle }{\tau_{1}-\tau_{2}},\quad\zeta_{2}&=\tau_{2}\frac{\left\langle \boldsymbol{\Pi},\boldsymbol{z}\right\rangle -\tau_{1}\left\langle \boldsymbol{\Pi},\Tmati\boldsymbol{z}\right\rangle }{\tau_{2}-\tau_{1}}.
\end{align}
\end{ex}

Figure \ref{fig:flowchart} shows a flowchart describing the general process of obtaining the evolution equation and/or the transport coefficients. It is important to highlight that for a fixed $n$, the approach of this work gives both the evolution equation as well as the transport coefficients in an explicit form composed of matrix invariants. Explicit examples in terms of matrix invariants can be found in the appendix for small $n$ (including, for completeness, a reiterated $n=2$ without specifying the particle and reaction content). 

\section{Discussion and outlook} \label{sec:discussion}

We have analysed the transport properties of a general dense system with $n$ independent chemical potentials in second-order hydrodynamics in the linearised regime. In particular, we found the evolution equation \eqref{eq:pieq} for the bulk scalar as well as the relevant transport coefficients (relaxation times and bulk viscosities) in two different representations, both expressed in terms of equilibrium rates, densities, and susceptibilities. In our analysis, we found that the system factorises into an $n$-component fluid, with each component evolving according to the Israel--Stewart equation.  

In particular, the generally convoluted expressions for the bulk viscosities for the subsystems are given by simple expressions (as long as the bulk scalar splits into $n$ components, e.g. in the Navier--Stokes regime) and by far the most complex part of their determination is essentially reduced to determining the eigenvalues of an $n\times n$-matrix $\Tmat$. 

The form obtained for the transport coefficients is readily implemented for a given physical system. This will be especially useful for e.g. numerical purposes in viscous hydrodynamical simulations in comparison to directly writing out the expressions in terms of susceptibilities, rates, and densities, which already at $n=2$ results in remarkably unwieldy expressions. 

The results found will be useful when studying the transport properties of increasingly complicated systems involving multiple particle species, which arise naturally in the multi-faceted setting of neutron star physics. As the results and techniques used have been rather general, they improve the general understanding of viscous systems and will hopefully prove helpful for the study of viscosity in systems beyond neutron stars, as well.

\section{Acknowledgements}
I acknowledge support from the program Unidad de Excelencia María de Maeztu CEX2020-001058-M, from the
project PID2022-139427NB-I00 financed by the Spanish MCIN/AEI/10.13039 /501100011033/FEDER, UE (FSE+), as well as from the Generalitat de Catalunya under contract 2021 SGR 171.  I also thank Cristina Manuel for not only interesting discussions during the preparation of \cite{Hernandez:2025zxw} which led me to investigate the $n$-component generalisation, but also for encouraging me to write the generalisation down, as well as both C. Manuel and Aleksi Vuorinen for valuable feedback on an initial version of this work.

\appendix

\section{Explicit results for small $n$} \label{sec:appendix}

For a fixed $n$, the coefficients $c_i[\Tmati]$ appearing in the characteristic polynomial can be written down in terms of matrix invariants, e.g. with the Faddeev--LeVerrier algorithm \cite{Hou:1998}.  Here, we reproduce such forms of the evolution equation and form of the coefficients of \eqref{eq:pieq} for $n=2$, $n=3$ which have been studied in the context of compact starts, and also display $n=4$. Regardless of the applications cited here, we emphasise that the method is general for systems following differential equations of this form. For $n=1,2$, we reproduce results such as \cite{Hernandez:2025zxw,CruzRojas:2024etx,Hernandez:2024rxi,Gavassino:2023eoz} and for $n=3$ we find a considerable simplification to the forms presented in \cite{Harris:2024ssp}. As was already seen in \cite{Hernandez:2025zxw} (see the appendix), the simplification using linear algebra is considerable over working with the explicit coefficients of the matrices. This process is feasible for any $n$, even when explicitly solving the eigenvalues is generally not possible. 

\subsection{$n=2$}
The results for $n=2$ can be collected from the examples in the main text, but we list them all below. From $p_{\Tmati}(x)=x^2- \tr \Tmati x +\det \Tmati $ the evolution equation \eqref{eq:pieq} becomes
\begin{align}
\det \Tmat D_u^2 \Pi &+ \det \Tmat \tr \Tmati D_u \Pi + \Pi \nonumber \\
&=- \langle \boldsymbol{\Pi}, \bvec \rangle \vartheta - \det \Tmat \langle \boldsymbol{\Pi},\Tmati \bvec \rangle D_u \vartheta .
\end{align}
For any $n$, the inverse relaxation times $\set{\tau_\alpha^{-1}}_{\alpha=1}^{n}$ are the eigenvalues of $\Tmati$. The bulk viscous coefficients are
\begin{align}
Z_1 = \zeta_1 + \zeta_2 = \langle \boldsymbol{\Pi},\bvec \rangle,\quad Z_2 = \zeta_1 \tau_2 + \zeta_2 \tau_1 =  \det \Tmat \langle \boldsymbol{\Pi},\Tmati \bvec \rangle
\end{align}
or alternatively
\begin{align}
\zeta_{1}=\frac{\tau_{1}Z_{1}-Z_{2}}{\tau_{1}-\tau_{2}}=\tau_1\frac{\left\langle \boldsymbol{\Pi},\boldsymbol{z}\right\rangle -\tau_{2}\left\langle \boldsymbol{\Pi},\Tmati\boldsymbol{z}\right\rangle }{\tau_{1}-\tau_{2}},\quad\zeta_{2}=\frac{\tau_{2}Z_{1}-Z_{2}}{\tau_{2}-\tau_{1}}=\tau_{2}\frac{\left\langle \boldsymbol{\Pi},\boldsymbol{z}\right\rangle -\tau_{1}\left\langle \boldsymbol{\Pi},\Tmati\boldsymbol{z}\right\rangle }{\tau_{2}-\tau_{1}}.
\end{align}

\subsection{$n=3$}
For $n=3$ the equations are still very reasonable: The evolution equation is 
\begin{align}
\det \Tmat D_{u}^{3}\Pi&+\det \Tmat \tr\Tmati D_{u}^{2}\Pi+\frac{\det \Tmat}{2}\left(\tr^{2}\Tmati-\tr\Tmati^{2}\right)D_{u}\Pi+\Pi\nonumber \\
&=-\langle \boldsymbol{\Pi} , \bvec \rangle\vartheta-\det \Tmat\left(\tr\Tmati\langle \boldsymbol{\Pi} , \Tmati \bvec \rangle-\langle \boldsymbol{\Pi} , \Tmati^2\bvec \rangle\right)D_{u}\vartheta- \det \Tmat\langle \boldsymbol{\Pi} , \Tmati \bvec \rangle D_u^2 \vartheta.
\end{align}
and the bulk viscous coefficients are
\begin{align}
Z_{1}&=\zeta_{1}+\zeta_{2}+\zeta_{3}=\left\langle \boldsymbol{\Pi},\boldsymbol{z}\right\rangle \nonumber, \\
Z_{2}&=\left(\tau_{2}+\tau_{3}\right)\zeta_{1}+\left(\tau_{1}+\tau_{3}\right)\zeta_{2}+\left(\tau_{1}+\tau_{2}\right)\zeta_{3}=\det\Tmat\left(\tr\Tmati\left\langle \boldsymbol{\Pi},\Tmati\boldsymbol{z}\right\rangle -\left\langle \boldsymbol{\Pi},\Tmati^{2}\boldsymbol{z}\right\rangle \right) \nonumber, \\
Z_{3}&=\tau_{2}\tau_{3}\zeta_{1}+\tau_{1}\tau_{3}\zeta_{2}+\tau_{1}\tau_{2}\zeta_{3}=\det\Tmat\left\langle \boldsymbol{\Pi},\Tmati\boldsymbol{z}\right\rangle.
\end{align}
As mentioned in the main text, for any $n>1$ all but one of the bulk viscosities reduces to a form found in the $n-1$-component system. The partial bulk viscosities read

\begin{align}
\zeta_{1}&=\frac{\tau_{1}^{2}\left(\left\langle \boldsymbol{\Pi},\boldsymbol{z}\right\rangle -\left(\tau_{2}+\tau_{3}\right)\left\langle \boldsymbol{\Pi},\Tmati\boldsymbol{z}\right\rangle +\tau_{2}\tau_{3}\left\langle \boldsymbol{\Pi},\Tmati^{2}\boldsymbol{z}\right\rangle \right)}{\left(\tau_{1}-\tau_{2}\right)\left(\tau_{1}-\tau_{3}\right)},\nonumber \\ 
\zeta_{2}&=\frac{\tau_{2}^{2}\left(\left\langle \boldsymbol{\Pi},\boldsymbol{z}\right\rangle -\left(\tau_{1}+\tau_{3}\right)\left\langle \boldsymbol{\Pi},\Tmati\boldsymbol{z}\right\rangle +\tau_{1}\tau_{3}\left\langle \boldsymbol{\Pi},\Tmati^{2}\boldsymbol{z}\right\rangle \right)}{\left(\tau_{2}-\tau_{1}\right)\left(\tau_{2}-\tau_{3}\right)},\nonumber \\
\zeta_{3}&=\frac{\tau_{3}^{2}\left(\left\langle \boldsymbol{\Pi},\boldsymbol{z}\right\rangle -\left(\tau_{1}+\tau_{2}\right)\left\langle \boldsymbol{\Pi},\Tmati\boldsymbol{z}\right\rangle +\tau_{1}\tau_{2}\left\langle \boldsymbol{\Pi},\Tmati^{2}\boldsymbol{z}\right\rangle \right)}{\left(\tau_{3}-\tau_{1}\right)\left(\tau_{3}-\tau_{2}\right)}.
\end{align}

\subsection{$n=4$}
Even for $n=4$, which to the best of our knowledge has not been considered in this context even in specific simplifying cases, the results are simple enough to be written down in a relatively compact form, although in practice it would be less tedious to do it indirectly using the coefficients of the characteristic polynomial. The evolution equation is 
\begin{align}
&\det\Tmat D_{u}^{4}\Pi+\det\Tmat\tr\Tmati D_{u}^{3}\Pi+\frac{\det\Tmat}{2}\left(\tr^{2}\Tmati-\tr\Tmati^{2}\right)D_{u}^{2}\Pi\nonumber\\
&\qquad+\frac{\det\Tmat}{6}\left(\tr^{3}\Tmati-3\tr^{2}\Tmati\tr\Tmati+2\tr\Tmati^{3}\right)+\Pi\nonumber\\
&=-\left\langle \boldsymbol{\Pi},\boldsymbol{z}\right\rangle \vartheta-\frac{\det\Tmat}{2}\Big[\left(\tr^{2}\Tmati-\tr\Tmati^{2}\right)\left\langle \boldsymbol{\Pi},\Tmati\boldsymbol{z}\right\rangle 
-2\tr\Tmati\left\langle \boldsymbol{\Pi},\Tmati^{2}\boldsymbol{z}\right\rangle +2\left\langle \boldsymbol{\Pi},\Tmati^{3}\boldsymbol{z}\right\rangle \Big]D_{u}\vartheta\nonumber\\
&\qquad-\det\Tmat\left(\tr\Tmati\left\langle \boldsymbol{\Pi},\Tmati\boldsymbol{z}\right\rangle -\left\langle \boldsymbol{\Pi},\Tmati^{2}\boldsymbol{z}\right\rangle \right)D_{u}^{2}\vartheta-\det\Tmat\left\langle \boldsymbol{\Pi},\Tmati\boldsymbol{z}\right\rangle D_{u}^{3}\vartheta,
\end{align}
with bulk viscous coefficients
\begin{align}
Z_{1}&=\left\langle \boldsymbol{\Pi},\boldsymbol{z}\right\rangle,\nonumber \\
Z_{2}&=\frac{\det\Tmat}{2}\left[\left(\tr^{2}\Tmati-\tr\Tmati^{2}\right)\left\langle \boldsymbol{\Pi},\Tmati\boldsymbol{z}\right\rangle -2\tr\Tmati\left\langle \boldsymbol{\Pi},\Tmati^{2}\boldsymbol{z}\right\rangle +2\left\langle \boldsymbol{\Pi},\Tmati^{3}\boldsymbol{z}\right\rangle \right],\nonumber \\
Z_{3}&=\det\Tmat\left(\tr\Tmati\left\langle \boldsymbol{\Pi},\Tmati\boldsymbol{z}\right\rangle -\left\langle \boldsymbol{\Pi},\Tmati^{2}\boldsymbol{z}\right\rangle \right),\nonumber \\
Z_{4}&=\det\Tmat\left\langle \boldsymbol{\Pi},\Tmati\boldsymbol{z}\right\rangle .
\end{align}
The partial bulk viscocities again follow the general pattern from Eq. \eqref{eq:zetasol}.

\bibliography{bibliography}
\end{document}